\documentclass[preprint,superscriptaddress,aps,prb,amsmath]{revtex4}
\usepackage{graphicx}
\usepackage{color}
\usepackage{amsmath}
\usepackage[american]{babel}

\begin{document}

\title{What limits supercurrents in high temperature superconductors?
A microscopic model of cuprate grain boundaries}

\author{S. Graser}
\email{siegfried.graser@physik.uni-augsburg.de}
\affiliation{Center for Electronic Correlations and Magnetism, Institute of Physics, 
University of Augsburg, D-86135 Augsburg, Germany}
\affiliation{Department of Physics, University of Florida, Gainesville, FL 32611, USA}

\author{P. J. Hirschfeld}
\affiliation{Department of Physics, University of Florida, Gainesville, FL 32611, USA}

\author{T. Kopp}
\affiliation{Center for Electronic Correlations and Magnetism, Institute of Physics, 
University of Augsburg, D-86135 Augsburg, Germany}

\author{R. Gutser}
\affiliation{Center for Electronic Correlations and Magnetism, Institute of Physics, 
University of Augsburg, D-86135 Augsburg, Germany}

\author{B. M. Andersen}
\affiliation{Niels Bohr Institute, University of Copenhagen,
Universitetsparken 5, DK-2100 Copenhagen, Denmark}

\author{J. Mannhart}
\affiliation{Center for Electronic Correlations and Magnetism, Institute of Physics, 
University of Augsburg, D-86135 Augsburg, Germany}

\renewcommand\abstractname{}
\begin{abstract}
\vspace{1.5cm}
{\bf The interface properties of high-temperature cuprate 
superconductors have been of interest for
many years, and play an essential role in Josephson junctions,
superconducting cables, and microwave electronics.  
In particular, the maximum critical current achievable in 
high-$T_c$ wires and tapes is well known to be limited by the 
presence of grain boundaries~\cite{Mannhart},
regions of mismatch between crystallites with misoriented
crystalline axes. In studies of single, artificially fabricated grain
boundaries the striking observation has been made that the
critical current $J_c$ of a grain boundary
junction depends {\em exponentially} on the misorientation 
angle~\cite{Dimosetal88}. Until now microscopic understanding of
this apparently universal behavior has been lacking.
We present here the results of a microscopic evaluation based on a
construction of fully 3D YBa$_2$Cu$_3$O$_{7-\delta}$~ grain boundaries by molecular
dynamics. With these structures, we calculate an 
effective tight-binding Hamiltonian for the 
$d$-wave superconductor with a grain boundary. 
The critical current is then shown to follow an
exponential suppression with grain boundary angle.
We identify the buildup of charge inhomogeneities 
as the dominant mechanism for the suppression of the
supercurrent.}
\end{abstract}

\date{\today}

\maketitle

To explain the exponential dependence of the critical current $J_c$ on 
the misorientation angle $\alpha$ Chaudhari and 
collaborators~\cite{Chaudhari} introduced several 
effects which can influence the critical current that are particular 
to high-temperature superconductor (HTS) grain boundaries (GB). 
First, a variation with angle can arise from the relative orientation
of the $d$-wave order parameters pinned to the crystal lattices
on either side of the boundary. This scenario was investigated 
in detail by Sigrist and Rice~\cite{SigristRice}. However, such 
a modelling cannot explain the exponential suppression of the 
critical current over the full range of misorientation angles.
Secondly, dislocation cores, whose density grows with increasing angle,
can suppress the total current. A model assuming insulating dislocation cores
which nucleate antiferromagnetic regions and destroy superconducting order 
was studied by Gurevich and Pashitskii~\cite{Gurevich}. However, for 
grain boundary angles beyond approximately 10$^\circ$ 
when the cores start to overlap this model fails. 
Finally, variations of the stoichiometry in the grain boundary region,
such as in the oxygen concentration, may affect the scattering of carriers
and consequently the critical current. Stolbov and collaborators~\cite{Stolbov} 
as well as Pennycook and collaborators~\cite{Pennycook} 
have examined the bond length distribution 
near the grain boundary and calculated the change in
the density of states at the Fermi level, 
or the change in the Cu valence, respectively.
In the latter work the authors used the reduced valences
to define an effective barrier near the boundary whose
width grows linearly with misorientation angle.

A critical examination of the existing models shows that
the difficulty of the longstanding HTS ``grain
boundary problem" arises from the multiple length
scales involved: atomic scale reconstruction of the interface, the
electrostatic screening length, the antiferromagnetic correlation
length, and the coherence length of the superconductor. Thus it
seems likely that only a multiscale approach to the problem can
succeed. 
 
Our goal in this paper is to simulate, in
the most realistic way possible, the nature of the actual grain
boundary in a cuprate HTS system, in order to
characterize the multiple scales which cause the exponential suppression
of the angle dependent critical current.
To achieve this goal we proceed in a stepwise fashion, first
simulating the atomic structure of realistic YBCO grain boundaries
and assuring ourselves that our simulations are robust and
duplicate the systematics of actual grain boundaries. Subsequently
we construct an effective disordered
tight-binding model, including $d$-wave pairing, whose parameters
depend on the structures of the simulated grain boundaries in a
well-defined way. Thus for any angle it will be possible to
calculate the critical current; then, for a given
pairing amplitude (reasonably well known from experiments on 
bulk systems) the form of $J_c(\alpha)$ and its absolute magnitude
is calculated.

\begin{figure}
\begin{centering}
\includegraphics[width=0.7\textwidth]{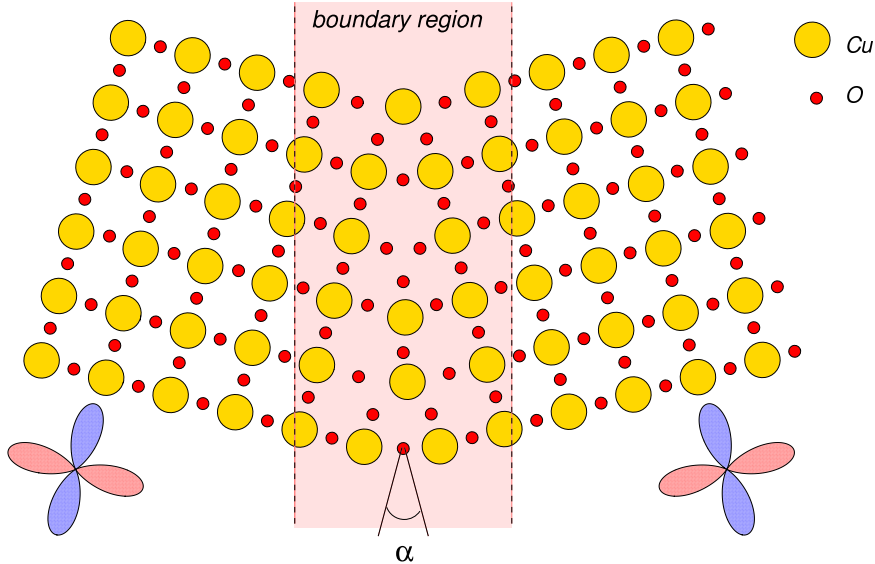}
\end{centering}
\caption{{\bf Schematic of an HTS symmetric grain boundary.} The
misorientation angle $\alpha$ and the orientation of the $d$-wave 
order parameters are indicated.} \label{schematic}
\end{figure}

We simulate YBCO grain boundaries by a molecular dynamics (MD)
procedure which has been shown to reproduce the correct structure
and lattice parameters of the bulk YBa$_2$Cu$_3$O$_{7-\delta}$~ 
crystal~\cite{Zhang}, and adapt techniques which were successfully 
applied to twist grain boundaries in monocomponent solids~\cite{Phillpot}.
We only sketch the procedure here, and postpone its details to the 
supplementary information and a longer publication. The method
uses an energy functional within the canonical ensemble
\begin{equation}
H=\frac{1}{2}\sum_{i=1}^N \sum_{\alpha=1}^3 m_i \dot{r}_{i,\alpha}^2 
+ \frac{1}{2}\sum_{i=1}^{N}\sum_{j=1,j\neq i}^{N}U(r_{ij}),
\label{MD} 
\end{equation}
where $m_i$ is the mass of the ion and $U(r_{ij})$ is the 
effective potential between ions, taken to be of the form
$U(r_{ij})=\Phi(r_{ij})+V(r_{ij})$.  Here $V(r)$ is the screened
Coulomb interaction $ V(r)=\pm e^{-\kappa
r}{Z^{2}e^{2}}/({4\pi\epsilon_{0}}{r})$ with screening length
$\kappa^{-1}$, and $\Phi(r)$ is a short range Buckingham potential
$\Phi(r)=A\exp(-r/\rho)-C/r^{6}$.   We take the parameters $A$,
$\rho$ and $C$ from Ref.~\onlinecite{Zhang}, and the initial
lattice constants are $a=3.82$~\AA, $b=3.89$~\AA 
~and $c=11.68$~\AA. To construct a grain boundary with
well defined misorientation angle, we must fix the ion
positions on both sides far from the boundary,
and ensure that we have a periodic lattice structure along the
grain boundary and also along the $c$-axis direction. Therefore we
apply periodic boundary conditions in the molecular dynamics
procedure in the direction parallel to the GB and also in the
$c$-axis direction. In the direction perpendicular to the GB only
atomic positions with a distance smaller than six lattice constants from the GB
are reconstructed. This method restricts our consideration to
grain boundary angles that allow a commensurate structure
parallel to the GB, e.g. angles 
$\alpha=2 \cdot \arctan[{N_{1}a}/({N_{2}b})],~ N_{1},N_{2}\in\mathbf{N}$
\cite{gbangles}.

\begin{figure}
\begin{centering}
\includegraphics[width=0.8\textwidth]{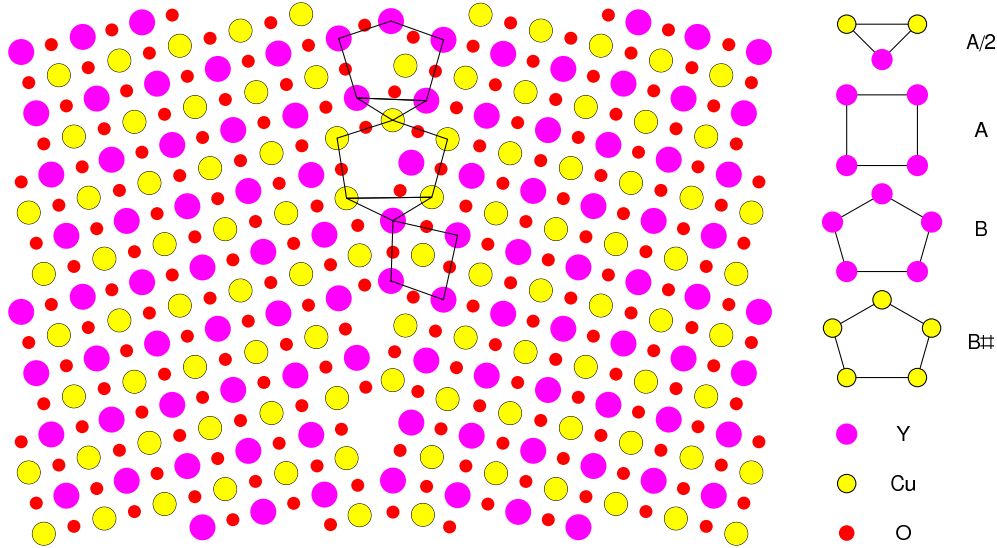}
\end{centering}
\caption{{\bf Top view of a calculated (410) grain boundary.} 
Here only the Y and the CuO$_2$ layers are shown. 
The dots show the position of
Y ions (magenta), Cu ions (yellow), and O ions
(red).  Structural units are indicated by solid black lines. For
this particular angle we find a sequence of the form
A(A/2)B\#(A/2)B, in agreement with the experimental results given
in Table 1 of Ref.~\onlinecite{Pennycook} (for notation see
this reference).} \label{410}
\end{figure}

An important step to be taken before starting the reconstruction of the GB is
the initial preparation of the GB. Since we use a fixed number of
ions at the GB within the molecular dynamics algorithm we have to
initialize it with the correct number of ions. If we start with two
perfect but rotated crystals on both sides of the interface, in
which all ions in the half space behind the
imaginary boundary line are cut away, we find that several
ions are unfavorably close to each other. Here we have to use a set
of selection rules to replace two ions by a single one at the
grain boundary. These selection rules have to be carefully chosen
for each type of ion since they determine how many ions of
each type are present in the grain boundary region; the rules are
detailed in the supplementary information. 
Ultimately they should be confirmed by a grand canonical MD procedure.
However, such a procedure for a complex multicomponent system as
the YBCO GB is technically still not feasible.
While the selection rules are
{\it ad hoc} in nature, we emphasize that they are
independent of misorientation angle, and reproduce very well the 
experimental TEM structures~\cite{Pennycook}.

With the initial conditions established, the MD equations of motion
associated with Eq.~\ref{MD} are iterated until all atoms are 
in equilibrium.  As an example,  we show in Fig.~\ref{410} the 
reconstructed positions of Y, Cu, and O ions for a (410) boundary.  
We emphasize that the MD
simulation is performed for {\it all} the ions in the 
YBCO full 3D unit cells of two misoriented crystals
except for a narrow ``frame" consisting of ions that are fixed to
preserve the crystalline order far from the grain boundary.  The
sequence of typical structural units identified in the experiments~\cite{Pennycook} 
is also indicated and we find excellent agreement.

\begin{figure}
\begin{centering}
\includegraphics[width=0.95\textwidth]{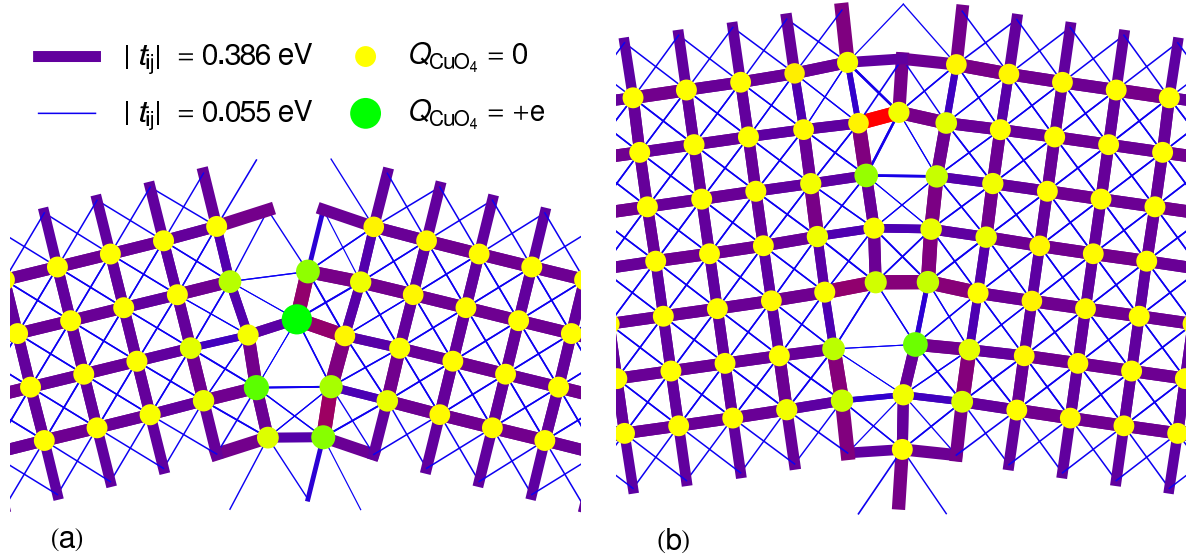}
\end{centering}
\caption{{\bf Tight-binding model for the CuO$_2$ plane.}
Hopping values between the Cu ions calculated from the 
interatomic matrix elements for a (410) grain boundary (a) 
and for a (710) grain boundary (b). The
line thickness shows the hopping amplitudes and the 
color and size of the copper sites illustrate the on-site potential.}
\label{tightbinding 410}
\end{figure}

We next proceed to construct an effective tight-binding model 
which is restricted to the Cu sites, the positions 
of which were determined through the algorithm described. We calculate hopping
matrix elements $t_{ij}$ of charge carriers (holes) up to next nearest
neighbor positions of Cu ions. 
The Slater-Koster method is used to calculate the directional dependent 
orbital overlaps of Cu-$3d$ and O-$2p$ orbitals~\cite{SlaterKoster,Harrison}. 
The effective hopping between Cu positions is a sum of direct orbital overlaps
and hopping via intermediate O sites, where the latter is calculated in 2nd order 
perturbation theory. For the homogeneous lattice, these parameters 
agree reasonably well with the numbers
typically used in the literature for YBCO. Exact values and
details of the procedure are given in the supplementary
information.  Results for a (410) grain boundary are shown in
Fig.~\ref{tightbinding 410}.  Note the largest hopping probabilities across the
grain boundary are associated with the 3fold coordinated Cu ions
which are close to dislocation cores. The inhomogeneities 
introduced through the distribution of hopping probabilities along the boundary
induce scattering processes of the charge carriers and consequently 
contribute to an ``effective barrier" at the grain boundary.
We note that the angular dependence of the critical current 
$J_c(\alpha)$ is not directly related to the changes in averaged 
hopping parameters observed for different misalignment angles.
In fact we found that the variation of the hopping
probabilities with boundary angle cannot account by itself for the 
exponential dependence of $J_c(\alpha)$ over the whole
range of misalignment angles.

\begin{figure}
\includegraphics[width=0.63\columnwidth]{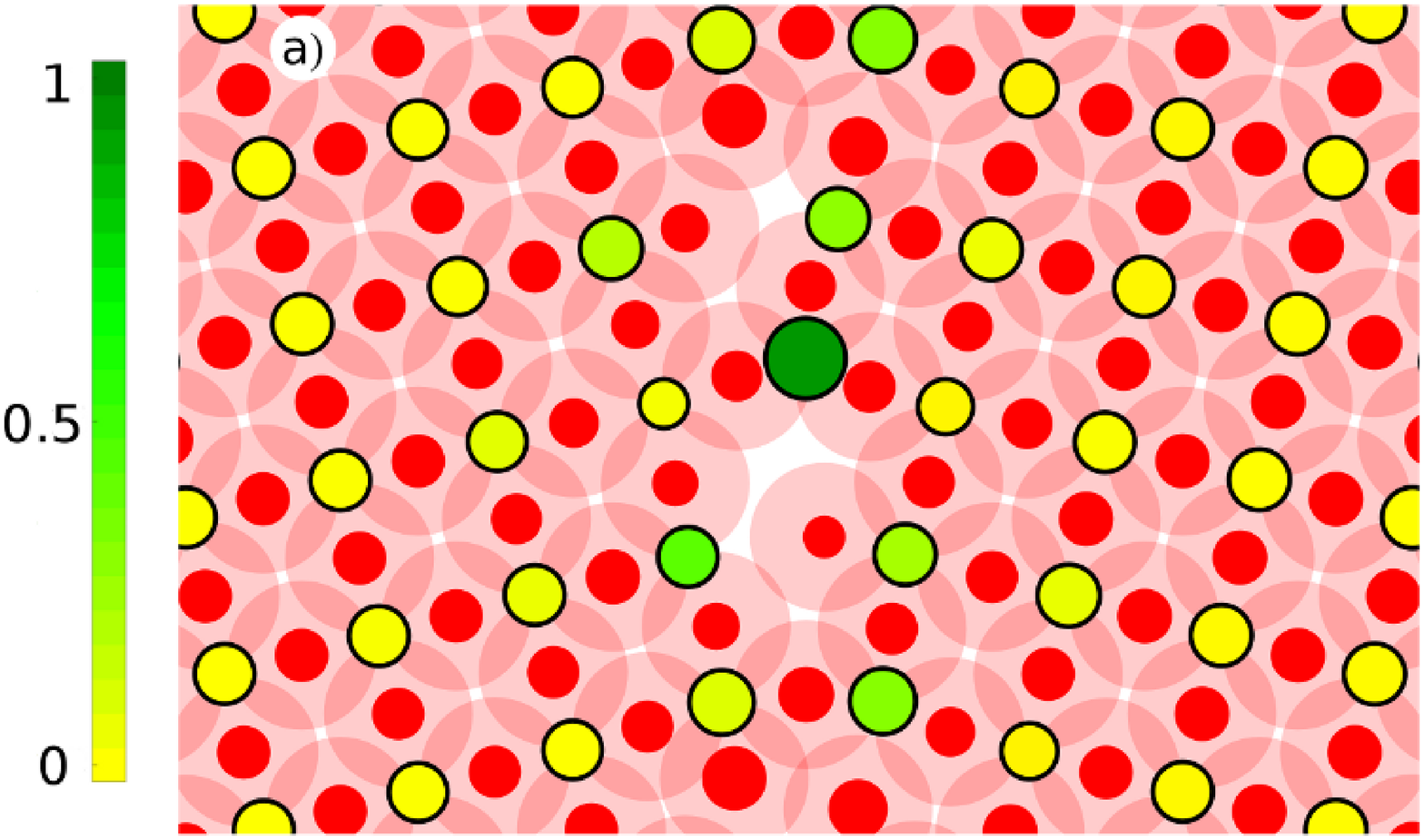}
\includegraphics[width=0.36\columnwidth]{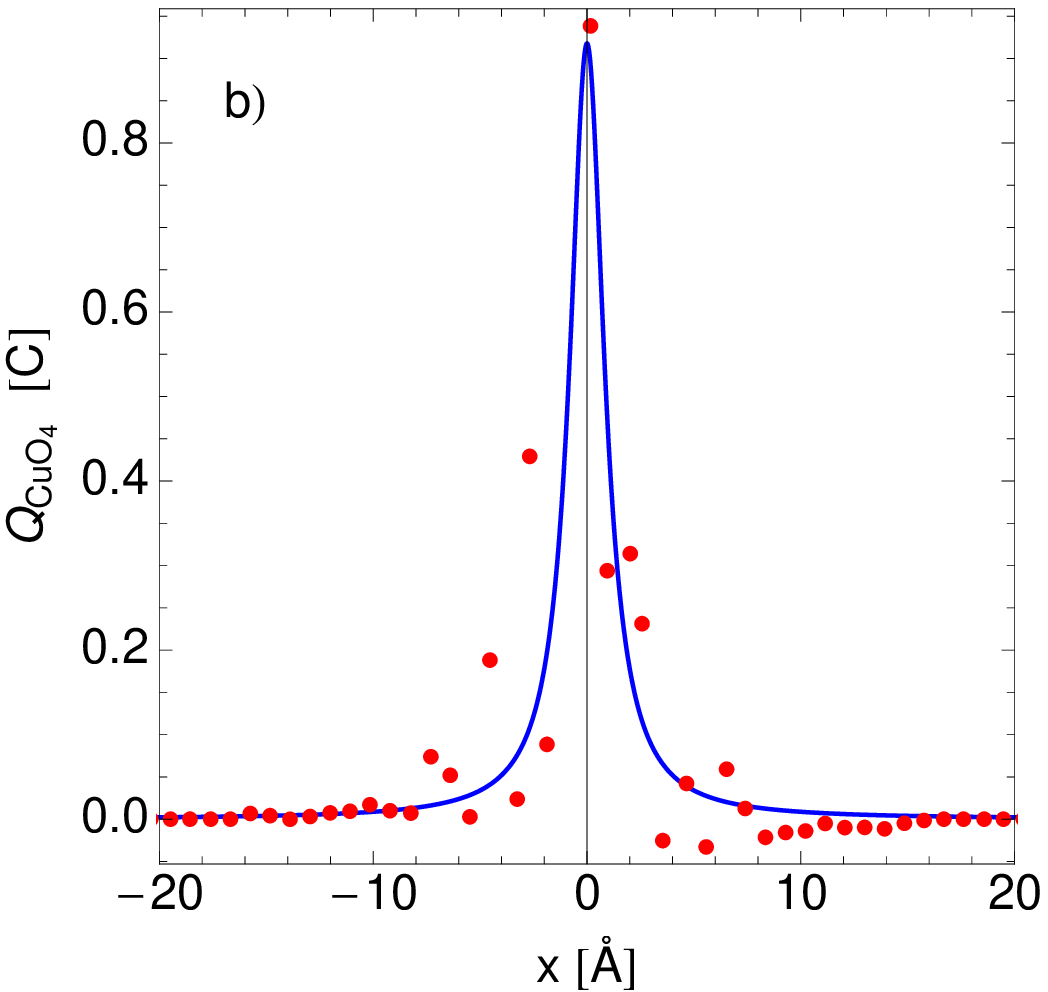}
\caption{{\bf Charging of the CuO$_4$ squares.}
a) Charge distribution on copper (yellow/green) and 
oxygen (red) sites at a 410 boundary. The diameter of the circles is a measure 
for positive (copper) or negative (oxygen) charge, as determined by the bond valence analysis.
The color of the copper sites represents the charging of the 
corresponding CuO$_4$ squares as described by 
Eq. \ref{CuO4charge}, with green circles referring
to a positive charge compared to the 
bulk charge (see color scale). The transparent red circles represent the 
oxygen contributions to the CuO$_4$ charge.
b) Plot of average charge on squares vs. distance from
grain boundary (red points), and fit by a Lorentzian (blue line).} \label{Cu_charges}
\end{figure}

The structural imperfection at the grain boundary will necessarily
lead to charge inhomogeneities that will contribute---in a similar
way as the reduced hopping probabilities---to the effective barrier that blocks
the superconducting current over the grain boundary. We
include these charge inhomogeneities into the calculations by
considering them in the effective Hamiltonian as on-site
potentials on the Cu sites. To accomplish this we utilize the
method of valence bond sums. The basic idea is to calculate the
bond valence of a cation by
\begin{equation}
V_i=\sum_j\exp\left(\frac{r_{0}-r_{ij}}{B}\right)
\end{equation} 
where $j$ runs over all neighboring anions, in our case the neighboring
negatively charged oxygen ions. The parameter $B=0.37$~\AA ~is a universal constant in the
bond valence theory, while $r_0$ is different for
all cation-anion pairs and also depends on the formal integer
oxidation state of the cation (the values are listed in Ref.~\onlinecite{Chmaissem}). 
Strong deviations from the formal valence reveal strain or even an incorrect
structure. This procedure is straightforward in the case of the Y$^{3+}$ and
Ba$^{2+}$ ions, while it is slightly more complicated for the Cu ions, because they 
have more than one formal integer oxidation state~\cite{Brown,Chmaissem}.  
We show in Fig.~\ref{Cu_charges} the distribution of charges at
the (410) grain boundary obtained. We also
calculate by similar methods the oxidation state of the oxygen
ions. Here, charge neutrality is ensured because the already determined
cation valences are used.

In the next step we account for the effect of broken Cu-O 
bonds at the grain boundary that give rise to strong changes 
of the electronic configuration of the Cu atoms as well as of 
the electronic screening of charges, as shown
in first principle calculations~\cite{Stolbov}. 
Unfortunately there is no straightforward way to include 
these changes in the electronic configuration 
into a purely Cu-based tight-binding Hamiltonian.
On the other hand we know that the additional holes doped into the 
CuO$_2$ planes form Zhang-Rice singlets residing on a CuO$_4$ square 
rather than on a single Cu site~\cite{ZhangRice}, and are therefore affected
not only by the charge of the Cu ion but also by the charge 
of the surrounding oxygens. Modelling this situation, we use 
a phenomenological potential to sum the Cu
and the O charges to obtain an effective charge of the
CuO$_{4}$ square. This effective charge is taken as
\begin{equation}
Q_{\textrm{CuO}_4}(i)=Q_{\textrm{Cu}}(i)+A\sum_jQ_0(j)e^{-r_{ij}^2/\lambda^2},
\label{CuO4charge}
\end{equation}
where $A$ and $\lambda$ are two constants chosen to yield
a neutral Cu site if 4 oxygen atoms are close to 
the average Cu-O distance. Correspondingly, the
energy cost of a Cu site that has only 3 close oxygen neighbors instead 
of all 4 neighbors, is strongly enhanced. Thus, the broken
Cu-O bonds induce strong charge inhomogeneities in the ``void'' regions
of the grain boundary, mainly described by pentagonal structural
units, while Cu sites belonging to ``bridge'' regions with
mostly quadrangular structural units have charge values close to
their bulk values.

Finally we have to translate the charge on the Cu sites (or better
CuO$_4$ squares) into effective on-site lattice potentials. The
values of screening lengths in the cuprates, and particularly near
grain boundaries, are not precisely known, but near optimal doping
they are of order of a lattice spacing or less.   We adopt the simplest
approach and assume a 3D Yukawa-type screening with
phenomenological length parameter $\ell$ of this order, and
find a potential integrated over a unit cell $V_0$ of 
\begin{equation}
\frac{V_0}{a^2 \bar{q}}\approx 4\pi
~(a_0 \ell)~\mathrm{Ryd}\approx 10~\mathrm{eV},
\end{equation} 
where $\bar{q}$ is the charge in units of the elementary charge, 
$a_{0}$ is the Bohr radius, and $\ell$ is taken to be $2$~\AA ~while
$a=4$~\AA.  Thus we find a surplus charge of a single
elementary charge integrated over a unit cell to produce an
effective potential of around $10$~eV. In the following we will use
an effective potential of either 6 or 10 eV, reflecting the
uncertainty in this parameter. The value of the effective potential
will affect the scale of the final critical current.

To calculate the order parameter profile and the current across the
grain boundary we  solve the Bogoliubov-de Gennes mean field
equations of inhomogeneous superconductivity self-consistently.
The Hamiltonian is 
\begin{equation}
\hat{H}=\sum_i\epsilon_i\hat{n}_{i\sigma}-\sum_{ij\sigma}t_{ij}
\hat{c}_{i\sigma}^{\dagger}\hat{c}_{j\sigma}+\sum_{ij}\left(\Delta_{ij}
\hat{c}_{i\uparrow}^{\dagger}\hat{c}_{j\downarrow}^{\dagger}+\mathrm{h.c.}\right),
\label{BdGHamiltonian}
\end{equation}
where the effective hopping parameters $t_{ij}$ are determined for a given
grain boundary by the procedure described above and the onsite energies 
$\epsilon_i=u_i-\mu$ are a sum of the effective charge potentials $u_i$ 
and the chemical potential $\mu$. Performing a
Bogoliubov transformation, we find equations for the particle and
hole amplitudes $u_n$ and $v_n$
\begin{equation}
\sum_j\left(\begin{array}{cc}
H_{ij} & \Delta_{ij}\\
{\Delta}_{ij}^* &
-H_{ij}\end{array}\right)\left(\begin{array}{c}
u_n(j)\\
v_n(j)\end{array}\right)=E_n\left(\begin{array}{c}
u_n(i)\\
v_n(i)\end{array}\right)
\end{equation} 
with
\begin{equation}
H_{ij}=\epsilon_i\delta_{ij}-t_{ij}
\end{equation}
The self-consistency
equation for the $d$-wave order parameter is then
\begin{equation}
\Delta_{ij}=\frac{V_{ij}}{2 N_{\textrm{sc}}}\sum_{k_y}
\sum_n \left[ u_n(r_i)v_n^*(r_j) f(-E_n) -v_n^*(r_i)u_n(r_j)f(E_n) \right],
\end{equation}
where we use $N_{\textrm{sc}}$ supercells in the direction parallel to the
GB and $k_y$ is the corresponding Bloch wave vector. 
We adjust the chemical potential $\mu$ to ensure a fixed
carrier density in the superconducting leads corresponding to 
15$\%$ hole doping. 
The definition of the $d$-wave pair potential $V_{ij}$ 
in the vicinity of the grain boundary is not straightforward 
since the bonds connecting 
a Cu to its neighbors are not exactly oriented perpendicular 
to each other. We use a model that ties the strength of the 
pairing on a given bond to the size 
of the hopping on the bond, as well as to the charge difference 
across it, as detailed in the supplementary information. 
The final results for the critical 
current are not sensitive to the exact model employed.

\begin{figure}
\begin{centering}
\includegraphics[width=0.95\textwidth]{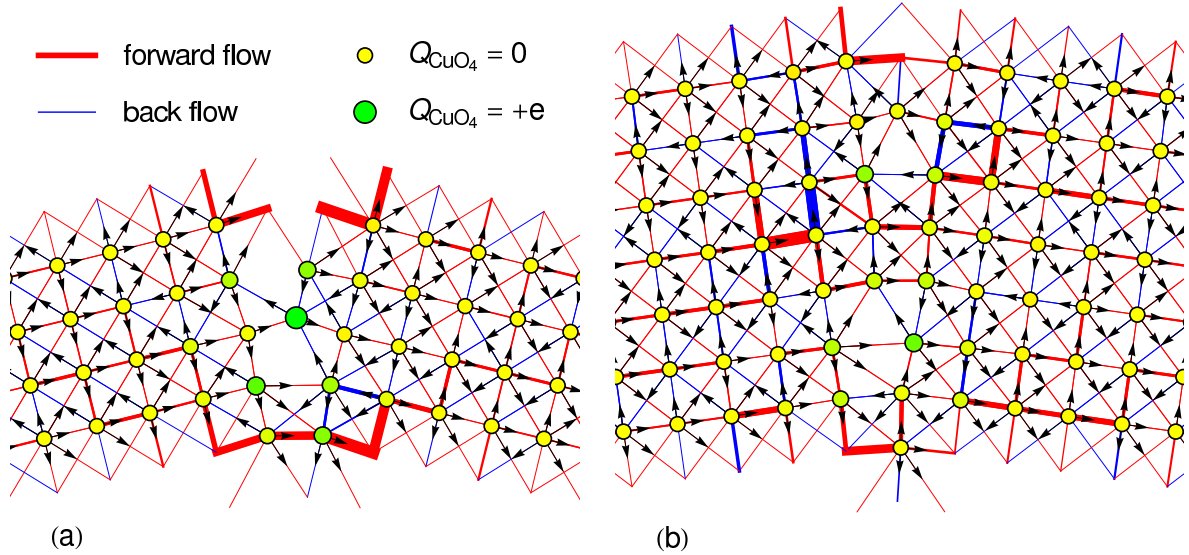}
\end{centering}

\caption{{\bf Supercurrent distribution.} 
The current pattern in the vicinity of a (410) (a) and
a (710) GB (b). The arrows only display the direction of the current, 
the red lines denote current flowing from left to right, blue lines denote
current from right to left. The line thickness shows the current
strength, while the point size and color of the Cu sites correspond to the 
on-site potential.}
\label{current}
\end{figure}

With these preliminaries, the current itself can finally be
calculated by imposing a phase gradient across the sample (see,
e.g. Ref.~\onlinecite{AndersenJc} for details) from the
eigenfunctions $u_n,v_n$ and eigenvalues $E_n$ of the BdG
equations for the grain boundary,
\begin{equation}
\frac{j(r_i,r_j)}{e/\hbar} = -\frac{2it_{ij}}{N_\textrm{sc}}
\sum_{k_y} \sum_n \left[u_n^*(r_i)u_n(r_j)
f(E_n)+v_n(r_i)v_n^* (r_j)f(-E_n) -\mathrm{h.c.} \right]
\end{equation}
The critical current $J_c$ is defined as the maximum value of the
current as a function of the phase.  In the tunnelling limit, when the
barrier is large, this relationship is sinusoidal so the maximum
current occurs at phase $\pi/2$.  However, for very low angle grain
boundaries we observe deviations from the tunnelling limit, i.e.
higher transparency grain boundaries with non sinusoidal
current-phase characteristics, although for the parameters studied here 
these deviations are rather small. 

It is instructive to examine the spatial pattern of supercurrent
flow across a grain boundary, which is far from simple, as
illustrated in Fig. \ref{current}. Along many bonds even away from
the boundary, the current flows backwards or 
runs in closed loops around the squares. The
flow appears to be dominated by large contributions between the
regions which resemble classical dislocation
cores.  In most of our simulated grain boundaries we do not
observe true  $\pi$ junction behavior, characterized
by an overall negative critical current. 
To derive the total current density across a 2D cross
section parallel to the grain boundary at $x=0$ we sum up all
contributions of $j(r_i,r_j)$ for which $x_i>0$ and
$x_j<0$, with $x=0$ the $x$ coordinate (perpendicular to the interface)
of the boundary, and normalize by the period length of the grain boundary
structure $ p=a/\sin\alpha$.

\begin{figure}
\begin{centering}
\includegraphics[width=0.95\textwidth]{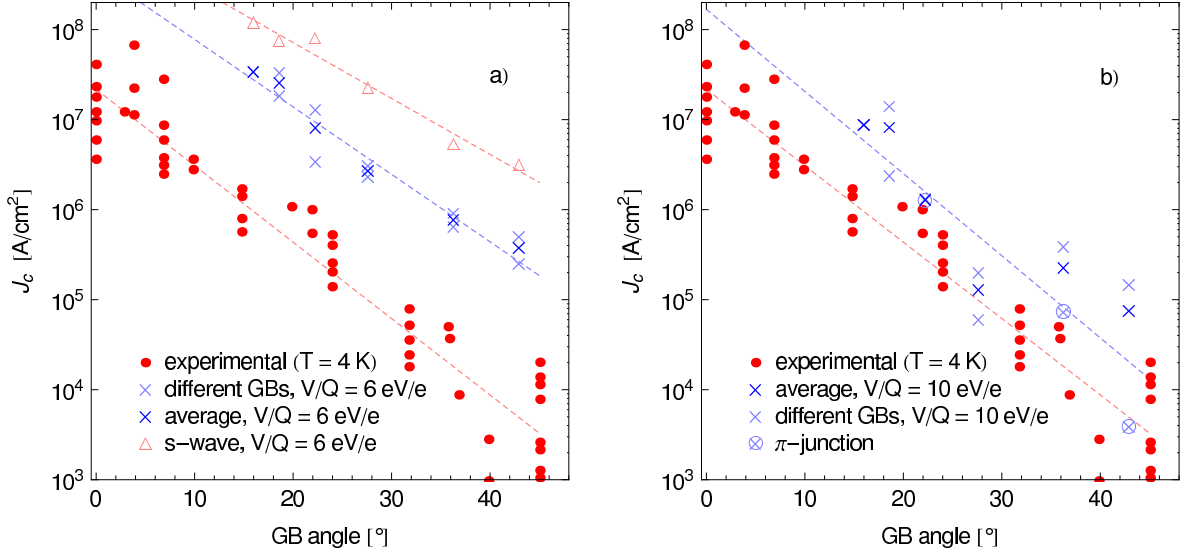}
\end{centering}

\caption{{\bf Angle dependence of the critical current.}
The critical current $J_c$  as a function of the grain
boundary angle $\alpha$ for screening
lengths $\ell = 1.2$~\AA~(a) and 2~\AA~(b). Here the red
points denote experimental results for YBCO junctions taken from 
Ref.~\onlinecite{Mannhart}, the light
blue crosses show theoretical results for differently reconstructed
GBs, the dark blue crosses show averaged theoretical values,
and the light red triangles show ``hypothetical'' $s$-wave results.
The dashed red and blue lines are exponential fits to the experimental 
and theoretical data, respectively.}
\label{jc}
\end{figure}

This calculation is in principle capable of providing the
absolute value of the critical current.  To accomplish this, we
have to normalize the current per grain boundary length by
the height of the crystal unit cell $c$ and
multiply it by the number of CuO planes per unit cell $N_\textrm{UC}$,
e.g. for the YBCO compound under consideration $c=11.7$~\AA, $N_\textrm{UC}=2$, 
\begin{equation}
j_{c}(x_{0})=\frac{N_\textrm{UC}}{pc}\sum_{i<j,x_i<0<x_j}j(r_i,r_j) 
\end{equation}
To account for the difference of the calculated gap magnitude from
its experimental value we multiply the current by a factor of 
$\Delta_\textrm{exp}/\Delta_0$, where $\Delta_\textrm{exp}$ is the experimentally 
measured order parameter and $\Delta_0$ is its self-consistently determined bulk
value.
We have checked that at low temperatures $T \ll T_c$ an approximately linear 
scaling of the critical current as a function of the order parameter 
holds true for all grain boundary angles.

In Fig. \ref{jc}, the critical current is plotted as a function
of misorientation angle for a set of grain boundary junctions from
(710) to (520) which we have simulated. All model parameters
are fixed for the different junctions, except for a range of
values affecting the initial conditions of the grain boundary
reconstruction, which resulted in slightly different structures
with the same misorientation angle $\alpha$.  Intriguingly, the
variability of our simulated junctions is quite similar to the
variability of actual physical samples as plotted in Fig.
\ref{jc}.  For two different choices of the screening length
$\ell$, we see that the dependence on misorientation angle
is exponential.  Since in our picture this parameter directly
affects the strength of the barrier, it is also natural that it
should affect the exponential decay constant, as also shown by the
Figure.  The value of $\ell$ which gives the correct slope
of the log plot yields a critical current which exceeds the experimental
value by an order of magnitude.  We speculate that the effect of
strong correlations (see Ref.~\onlinecite{Freericks} and references therein), 
not yet included in this theory, may account
for this discrepancy, given that a suppression by an order of
magnitude was already shown for (110) junctions\cite{Andersen_AF_SC}.  
We show in addition results for ``hypothetical'' $s$-wave junctions using
the same model parameters as for the $d$-wave junctions. We simply replaced
the bond-centered pair potential by an on-site pair potential 
resulting in an isotropic $s$-wave state. Although it is one order of 
magnitude larger, the critical current for the $s$-wave junction
still shows a similar exponential dependence on the grain boundary angle.
We emphasize that this model does not reflect the situation in 
a ``real'' $s$-wave superconductor like niobium or lead, that do not 
show an exponential angle dependence of the grain boundary current,
since it is based on the microscopic structures of a CuO$_2$ plane.
  
Our multiscale analysis of the grain boundary problem of HTS suggests
that the primary cause of the exponential dependence on
misorientation angle is the charging of the interface near defects
which resemble classical dislocation cores\cite{Gurevich}, leading
to a porous barrier where weak links are distributed in a
characteristic way which depends on the global characteristics of
the interface at a given angle (structure of defects, density of
dislocations, etc.). The $d$-wave order parameter  symmetry and the
nature of the atomic wave functions at the boundary which modulate
the hopping amplitudes do not appear to be essential
for the functional form of the angle dependence although they 
cannot be neglected in a quantitative analysis.  As such, we
predict that  this type of behavior may be observed in other
classes of complex superconducting materials.  Very recently, a
report of similar tendencies in ferropnictide grain boundary
junctions appeared to confirm this\cite{Larbalestier}.  It will be
interesting to use the new perspective on the longstanding
problem to try to understand how Ca doping of the grain boundaries
is able to increase the critical current by large
amounts\cite{Hammerl}, and to explore other chemical and
structural methods of accomplishing the same goal.

\begin{acknowledgments}
This work was supported by DOE grant DE-FG02-05ER46236 (PJH),
and by the DFG through SFB 484 (SG, TK, RG, and JM) and a research 
scholarship (SG). We are grateful to Y. Barash for important early
contributions to the project and we acknowledge fruitful 
discussions with A. Gurevich and F. Loder.  PH would also 
like to thank the Kavli Institute for Theoretical Physics 
for support under NSF-PHY05-51164 during the
writing of this manuscript.  The authors acknowledge the
University of Florida High-Performance Computing Center for
providing computational resources and support that have
contributed to the research results reported in this paper. 
\end{acknowledgments}

\newpage
\appendix
\section{Grain boundary reconstruction using molecular dynamics techniques}

The first step in our multiscale approach to determine the critical
current over a realistic tilt grain boundary (GB) is the modelling of the
microscopically disordered region in the vicinity of the seam of the two
rotated half crystals. Since the disorder introduced by the mismatch of the
two rotated lattices extends far into the leads on both sides of the GB plane,
simulations have to include a large number of atoms and it is very difficult
to perform them with high precision {\it ab initio} methods as for example density
function theory (DFT)~\cite{Schuster}. Here we have employed a molecular
dynamics technique to simulate the reconstruction of the ionic positions in
the vicinity of the GB starting from an initial setup containing
a realistic stoichiometry of each of the different ions. In the following we will
present the details of the calculational scheme.

\subsection{The initial setup of the grain boundary}

\begin{figure}
\begin{center}
\includegraphics[width=0.7\textwidth]{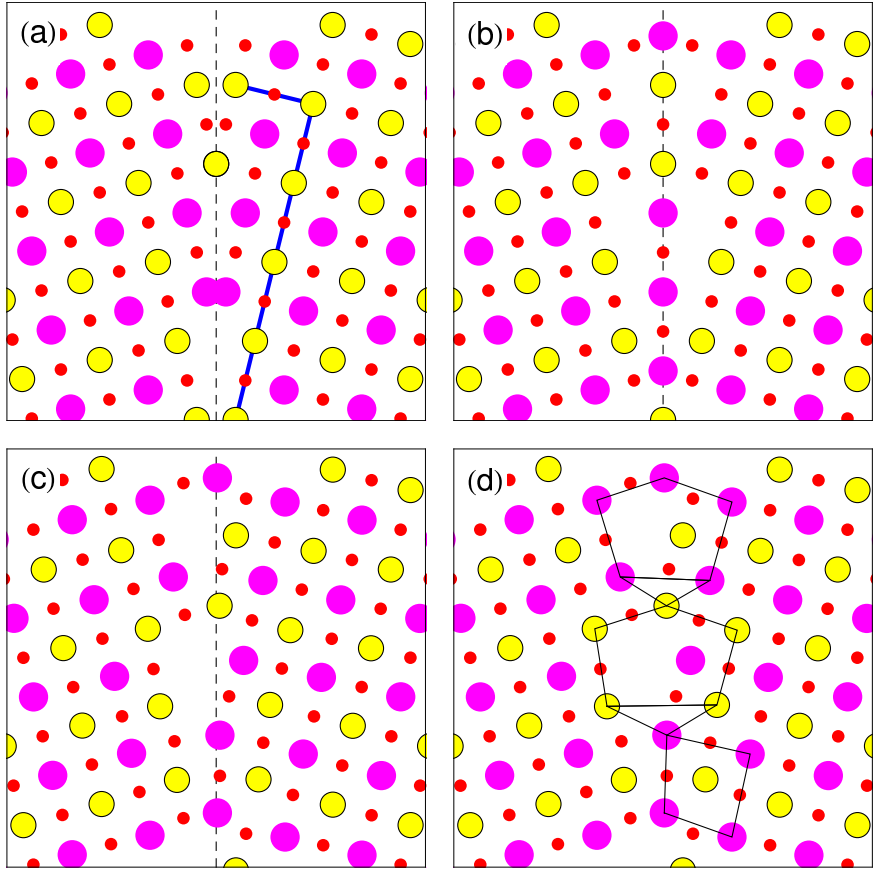}
\end{center}
\caption{The different steps in the reconstruction of a symmetric (410) GB:
(a)~Two half crystals rotated and cut behind the virtual GB plane (dashed line),
(b)~Initial setup of the GB using the ``selection rules'' outlined in the text,
(c)~Reconstructed GB using molecular dynamics,
(d)~Identification of basic structural units\cite{Pennycook}.
The blue line in (a) visualizes the classification of the ($mn0$)
GB with $m=4$ and $n=1$.
} \label{GBsetup}
\end{figure}

The simplest way of modelling a tilt grain boundary
is to stitch two perfect crystals together that are rotated
into one another around an axis in the plane of the GB perpendicular
to a lattice plane; atoms which cross into the region of space initially occupied by
the other crystal are then eliminated 
(see Fig.~\ref{GBsetup} a). If we examine the
structures constructed in this way we find on the one hand that some ions are left
unphysically close to each other, while on the other hand 
large ``void'' regions may also remain. Since a molecular dynamics scheme
employing an energy functional in the canonical ensemble does not allow
for the creation and annihilation of ions in the GB
region these faults in the setup will persist during the reconstruction process.
To improve the initial structures we develop a set of ``selection rules'':
(i) We introduce an overlap of the two crystals, extending them into a region
behind the virtual GB plane to prevent the creation of ``void'' regions.
(ii) We replace two ions that are too close to each other by a single
one at the GB.
Due to the different ionic radii, the different multiplicities and the
different internal positions these ``selection rules'' have to be
adjusted for every ion in the unit cell in order to reproduce the
correct stoichiometry found in TEM experiments~\cite{Pennycook}.
In Table~\ref{TabSelRules} we show the maximum overlap of ionic positions behind
the GB plane as well as the minimum distance between two ions allowed before
replacing them by a single ion at the GB. These values have proven to
reproduce the experimentally observed GB structures for all misorientation
angles examined within this work. An example of such a constructed GB
is shown in Fig.~\ref{GBsetup} b).

\begin{table}
\begin{center}
\begin{tabular}{|c||c|c|c|c|}
\hline
& Y$^{3+}$ & Ba$^{2+}$ & Cu$^{2+}$ (CuO$_2$) & Cu$^{2+}$ (CuO)\\
\hline
ov [$a$] & 0.15 & 0.2 & 0.15 & 0.15 \\
\hline
$d_\mathrm{min}$ [$a$] & 0.4 & 0.4 & 0.42 & 0.4 \\
\hline
\hline
& O$^{2-}$ (BaO) & O$^{2-}$ (CuO$_2$: $a$) & O$^{2-}$ (CuO$_2$: $b$) & O$^{-}$ (CuO) \\
\hline
ov [$a$] & 0.08 & 0.08 & 0.09 & 0.08 \\
\hline
$d_\mathrm{min}$ [$a$] & 0.3 & 0.35 & 0.3 & 0.3 \\
\hline
\end{tabular}
\caption{The maximum overlap behind the GB plane and the minimum distance
allowed between two ionic positions itemised by the different ion types.
The distances are given in units of the lattice constant $a = 3.82$~\AA.
\label{TabSelRules}}
\end{center}
\end{table}

\subsection{The molecular dynamics procedure}

\begin{figure}
\begin{center}
\includegraphics[width=0.4\textwidth]{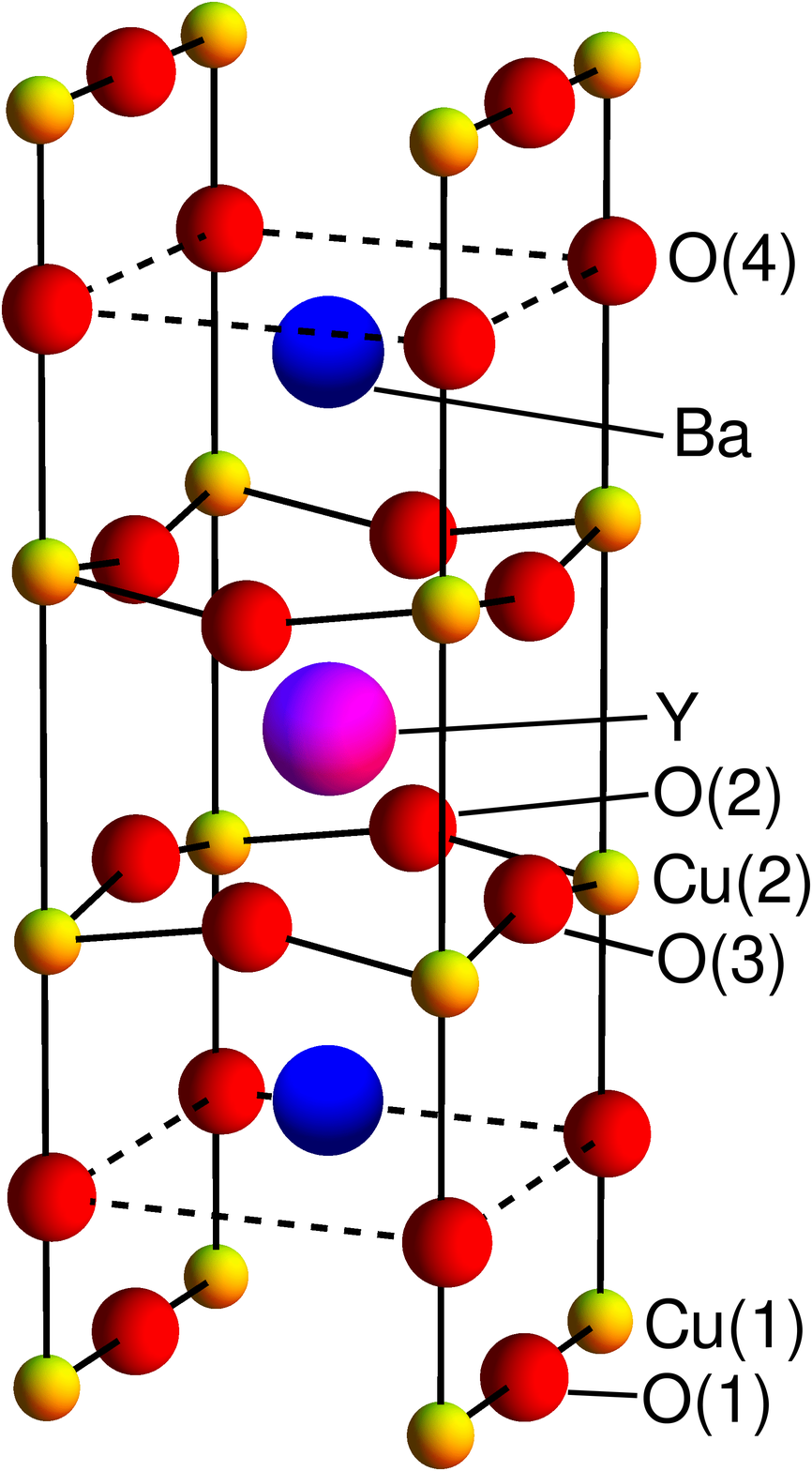}
\end{center}
\caption{The crystal structure found within the molecular dynamics
procedure calculated for a single unit cell with fixed lattice parameters
$a=3.82$~\AA, $b=3.89$~\AA, and $c=11.68$~\AA.} \label{crystal}
\end{figure}

The GB structures determined by applying the ``selection rules''
described in detail in the previous section are now used to initialize
the molecular dynamics process. For monocomponent
solids a method of zero temperature quenching has been successfully applied
for the reconstruction of high-angle twist GBs~\cite{Phillpot}.
Since this method uses an energy functional within the grand-canonical
ensemble it allows besides the movement of the atomic position
also the creation and annihilation of atoms.
For multicomponent systems like the complicated
perovskite-type structures of the high-$T_c$ superconductors this method is not
readily applicable. Therefore we choose a different approach using an energy
functional in the canonical ensemble with a fixed number of ions.
Here we can write the Lagrangian as
\begin{equation}
\mathcal{L} = \frac{1}{2}\sum_{i=1}^M \sum_{\alpha=1}^3 m_i \dot{r}_{i\alpha}^2
-\frac{1}{2}\sum_{i=1}^M \sum_{j=1,j\neq i}^M U(r_{ij}),
\end{equation}
and the Euler-Lagrange equations follow as the equations of motion for the ions
\begin{equation}
m_i \ddot{r}_{i,\alpha} = -\frac{1}{2}\sum_{j=1,j\neq i}^M
\frac{\partial U(r_{ij})}{\partial r_{i,\alpha}}.
\end{equation}
One of the main tasks is now the correct choice of
model potentials to ensure that the crystal structure of
YBa$_2$Cu$_3$O$_7$ (YBCO) is correctly reproduced for a homogeneous sample.
Here we use Born model potentials with long range
Coulomb interactions and short range terms of the Buckingham form
\begin{equation}
U(r_{ij})=\Phi(r_{ij})+V(r_{ij}).
\end{equation}
For the Coulomb interaction we can write the potential as
\begin{equation}
V(r)=\pm e^{-\kappa r}\frac{1}{4\pi\epsilon_{0}}\frac{Z^{2}e^{2}}{r},
\end{equation}
where we have introduced a Yukawa-type cut-off with $\kappa=\frac{1}{3.4}$~\AA$^{-1}$
to avoid the necessity to balance the long range Coulomb potentials
of the different ionic charges by the introduction of a Madelung constant.
For the short range Buckingham terms
\begin{equation}
\Phi(r)=A\exp(-r/\rho)-C/r^6
\end{equation}
we take the parameters $A$, $\rho$ and $C$ from molecular dynamics
studies by Zhang and Catlow~\cite{Zhang} leading to a stable
YBCO lattice with reasonable internal coordinates of
each atom (see Table~\ref{TabBuckParam}). In addition we
use the lattice constants $a=3.82$~\AA, $b=3.89$~\AA, and $c=11.68$~\AA~ when
setting up the initial GB structure and we fix them in the leads far away
from the GB. The CuO chains in the CuO layer are directed
parallel to the $b$-axis direction (compare Fig.~\ref{crystal}).

\begin{table}
\begin{center}
\begin{tabular}[b]{|c||c|c|c|c|}
\hline
{\bf A} & $A$ (eV) & $\rho$ (\AA) & $C$ (eV \AA$^6$) \\
\hline
\hline
O$^{2-}$-O$^{2-}$ & $22764.3$ & $0.149$ & $25.0$ \\
\hline
O$^{2-}$-O$^{-}$ &  $22764.3$ & $0.149$ & $25.0$ \\
\hline
O$^{2-}$-Cu$^{2+}$ & $3799.3$ & $0.243$ & $0$ \\
\hline
O$^{2-}$-Ba$^{2+}$ & $3115.5$ & $0.33583$ & $0$ \\
\hline
O$^{2-}$-Y$^{3+}$ & $20717.5$ &  $0.24203$ & $0$ \\
\hline
O$^{-}$-O$^{-}$ &  $22764.3$ & $0.149$ & $25.0$ \\
\hline
O$^{-}$-Cu$^{2+}$ & $1861.6$ & $0.25263$ & $0$ \\
\hline
O$^{-}$-Ba$^{2+}$ & $29906.5$ & $0.27238$ & $0$ \\
\hline
Cu$^{2+}$-Ba$^{2+}$ & $168128.6$ & $0.22873$ & $0$ \\
\hline
Ba$^{2+}$-Ba$^{2+}$ & $2663.7$ & $0.25580$ & $0$ \\
\hline
\end{tabular}
\;\;\;
\begin{tabular}[b]{|c||c|c|c|}
\hline
 {\bf B} & $d_{\textrm{MD}}$ (\AA) & $d_{\textrm{exp}}$ (\AA)  \\
\hline
\hline
Cu(1)-O(1) & $1.955$ & $1.94$ \\
\hline
Cu(1)-O(4) & $1.783$ & $1.847$ \\
\hline
Cu(2)-O(2) & $1.951$ & $1.925$ \\
\hline
Cu(2)-O(3) & $1.98$ & $1.957$ \\
\hline
Cu(2)-O(4) & $2.367$ & $2.299$ \\
\hline
Ba-O(1) & $3.058$ & $2.964$ \\
\hline
Ba-O(2) & $2.837$ & $2.944$ \\
\hline
Ba-O(3) & $2.797$ & $2.883$ \\
\hline
Ba-O(4) & $2.759$ & $2.740$ \\
\hline
Y-O(2) & $2.372$ & $2.407$ \\
\hline
Y-O(3) & $2.351$ & $2.381$ \\
\hline
\end{tabular}
\caption{(A) The parameters used to model the short range
potentials of the Buckingham form~\cite{Zhang}. (B) The bond lengths
found within the molecular dynamics ($d_{\textrm{MD}}$) compared to the experimental
values ($d_{\textrm{exp}}$)~\cite{Zhang}. The ions are labelled according to Fig.~\ref{crystal}.
\label{TabBuckParam}}
\end{center}
\end{table}

To construct a GB with well defined misalignment
angle we have to fix the atomic positions on both sides of the
interface. In addition we apply periodic boundary conditions
in the molecular dynamics procedure in the directions parallel to
the GB plane. In the direction perpendicular
to the GB only atoms with a distance from the GB plane
smaller than 6 lattice constants are reconstructed.

Since we are only interested in deriving stable equilibrium
positions for all atoms in the GB region and do
not try to simulate the temperature dependent dynamics of
the system we completely remove the kinetic energy at the end
of every iteration step. With this method the ions relax to
their equilibrium positions following paths given by classical
forces. With this procedure we are very likely to end up with
an ionic distribution that corresponds to a local minimum of the
potential energy instead of reaching the true ground state of
the system. Randomly changing the initial setup of the GB before
starting the reconstruction we are thus able to find different
GB structures corresponding to the same misalignment
angle $\alpha$. This reflects the experimental situation where one also
observes different patterns of ionic arrangements along a macroscopic grain
boundary with fixed misalignment angle. For all GB angles under
consideration (except the 710 GB) we have reconstructed and analyzed
two differently reconstructed grain boundary structures.
An example of a reconstructed 410 GB is shown in Fig.~\ref{GBsetup} c).
Finally we can identify the charateristic structural units
as classified in Ref.~\onlinecite{Pennycook}. Here we distinguish between
structural units of the bulk material and structural units
that are formed due to the lattice mismatch at the GB.
The first group consists of (deformed) rectangular
and triangular units, that can be
seen as fragments of a full rectangular unit, while the
latter group consists of large pentagonal units
bordered by either Cu or Y ions, that introduce strong
deformations and can be identified as the centres of classical
dislocation cores (see Fig.~\ref{GBsetup} d).

\section{The effective tight-binding model Hamiltonian}

\subsection{Slater-Koster method for the calculation of hopping matrix elements}

In the following we will derive a tight-binding
Hamiltonian for the CuO$_2$ planes with charge carriers located in the
$d_{x^{2}-y^{2}}$ orbitals of the copper atoms. The kinetic energy
associated with the hopping of charge carriers from one Cu site to one of its
neighboring sites can be calculated from the orbital overlaps of two
Cu-$d$ orbitals. Besides the direct overlap between two Cu-$d$ orbitals,
that is small due to the small spatial extension of the Cu-$3d$ orbitals,
we will also include the indirect hopping ``bridged'' by an O-$p$ orbital,
that can be calculated in second order perturbation theory.
Here we will have to add up all possible second order processes
involving the O-$p_x$ and O-$p_y$ orbitals of all intermediate
oxygen atoms.
In the vicinity of the grain boundary, the directional dependences of the
orbital overlaps become important and we calculate the interatomic hopping elements
from the Slater and Koster table of the displacement dependent interatomic
matrix elements~\cite{SlaterKoster,Harrison} that
depend on the direction cosines $l$, $m$ and $n$ of the vector
pointing from one  atom to the next, $\vec{r}=(l\vec{e}_{x}+m\vec{e}_{y}+n\vec{e}_{z})d$.
In addition, we calculate the effective potentials $V_{pd\sigma}$ and $V_{pd\pi}$ for
the $\sigma$- or $\pi$-bonds between the O-$p$ and the Cu-$d$ orbitals, as well as the potentials
$V_{dd\sigma}$ and $V_{dd\delta}$ for the $\sigma$- or $\delta$-bonds between
two Cu-$d$ orbitals using the effective parameters provided in Ref.~\onlinecite{Harrison}.
In the following we will outline the calculational scheme used within
this work by deriving the effective hopping parameter between two
neighboring Cu ions in a bulk configuration of a
flat CuO$_2$ plane with an average Cu-O distance of
\[
d_{\textrm{Cu-O}} = 1.95 \textrm{\AA} = 3.685 a_0.
\]
As a first step we calculate the hopping between the Cu-$d_{x^2-y^2}$ orbital and
the O-$p_x$ orbital that are connected by the vector $\vec{r}=d\vec{e}_x$ 
and therefore $l=1$ and $m=n=0$. The angular dependence is introduced as
\[
E_{x,x^2-y^2}=\frac{1}{2} 3^{1/2} l(l^2-m^2) V_{pd\sigma}
+l(1-l^2+m^2) V_{pd\pi} = \frac{1}{2} 3^{1/2} V_{pd\sigma}.
\]
In the next step we have to calcuate the distance-dependent potential of
the $\sigma$-bond
\[
V_{pd\sigma}=\eta_{pd\sigma} \frac{\hbar^2 r_d^{3/2}}{md^{7/2}}
= - 2.95 \cdot 7.62 \textrm{ eV \AA$^2$}
\frac{(0.67\textrm{\AA})^{3/2}}{(1.95\textrm{\AA})^{7/2}} = -1.19061\textrm{ eV},
\]
where we have used $\frac{\hbar^2}{m}=7.62$~eV \AA$^2$~
and the characteristic length $r_{d}=0.67$~\AA~ of the Cu-$d$ orbital has been taken
from Ref.~\onlinecite{Harrison}.
Now we can calculate the interatomic matrix element as
\[
E_{x,x^2-y^2}=\frac{1}{2} 3^{1/2} V_{pd\sigma} = -1.0311\textrm{ eV}.
\]
The corresponding hopping parameter between the Cu-$d$ and the O-$p_y$
orbital vanishes due to a basic symmetry argument.
The directional part of the direct overlap between two Cu-$d$ orbitals on 
nearest neighbor Cu sites ($d_{\textrm{Cu-Cu}}=3.9$~\AA)
can be calculated as
\[
E_{x^2-y^2,x^2-y^2}=\frac{3}{4} V_{dd\sigma} + \frac{1}{4} V_{dd\delta} = \frac{3}{4} V_{dd\sigma}.
\]
Again we need in addition the distance-dependent potential of the $\sigma$-bond
of two Cu-$d$ orbitals:
\[
V_{dd\sigma} = \eta_{dd\sigma} \frac{\hbar^2 r_d^3}{md^5}
= -16.2 \cdot 7.62 \textrm{ eV} \textrm{\AA}^2 \cdot
\frac{(0.67 \textrm{\AA})^3}{(3.9 \textrm{\AA})^5}=-0.04115\textrm{ eV},
\]
and the total energy associated with the direct overlap of two Cu-$d$ orbitals can
finally be calculated as
\[
E_{x^2-y^2,x^2-y^2}=\frac{3}{4} V_{dd\sigma}=-0.03086\textrm{ eV}.
\]
Now we can compare this energy to the kinetic energy describing
the superexchange between two Cu sites via the O-$p$ orbitals
\[
E_{\textrm{Cu-Cu}}= \frac{E_{x,x^2-y^2} \cdot E_{x,x^2-y^2}}{\epsilon_d-\epsilon_p}
+ \frac{E_{y,x^2-y^2}\cdot E_{y,x^2-y^2}}{\epsilon_d-\epsilon_p}
= \frac{1.06317}{-3.5} \textrm{ eV} + 0\textrm{ eV}=-0.304\textrm{ eV}.
\]
Due to a strong renormalization of the site energies
in the cuprates the effective charge transfer gap $\Delta=\epsilon_p-\epsilon_d$
is larger than one would expect from the difference of the bare site
energies of the Cu-$3d$ and the O-$2p$ orbitals~\cite{Alloul}.
Here we have chosen the charge transfer gap to be $\Delta=3.5$ eV,
a value that is consistent with the range of values found in
numerical studies. The full matrix element between
two Cu-$d$ orbitals is now the sum of the direct overlap and the second
order term including the intermediate O-$p$ orbitals.

\begin{figure}
\begin{center}
\includegraphics[width=\textwidth]{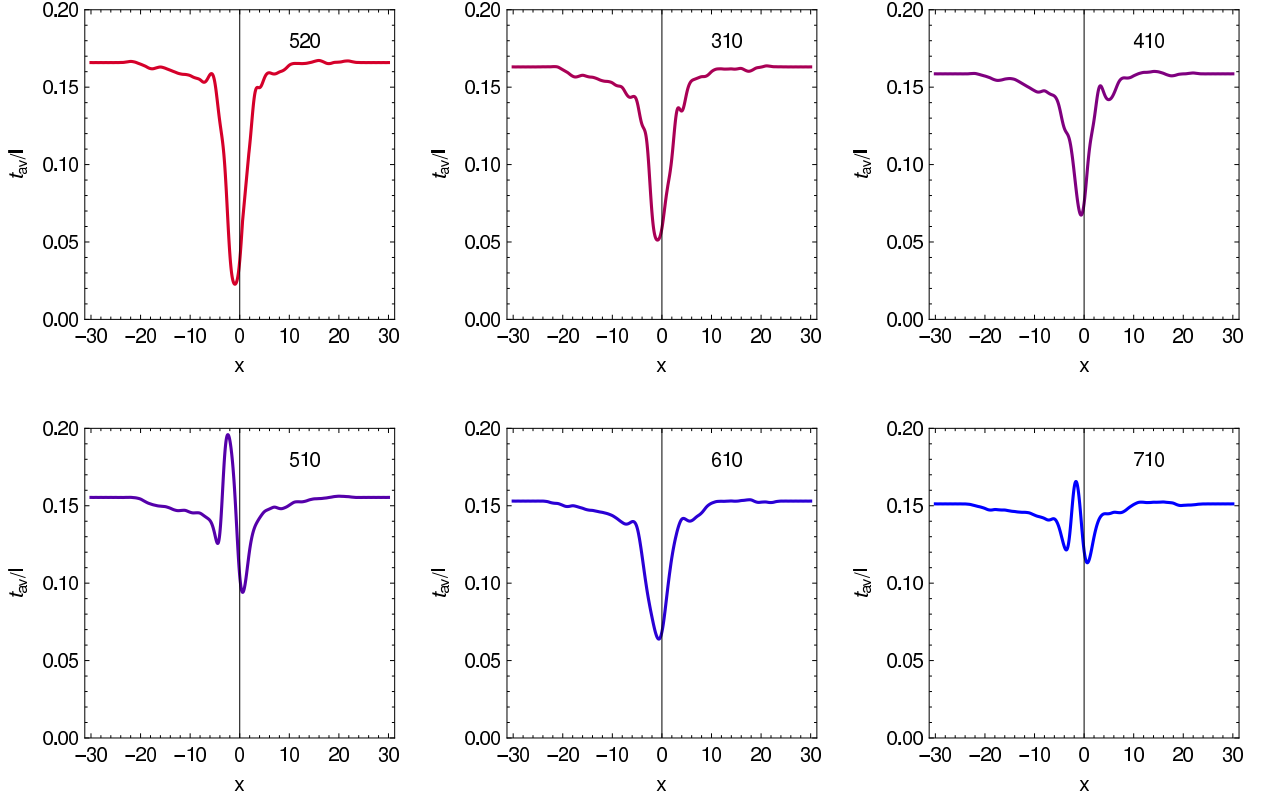}
\end{center}
\caption{The averaged hopping as a function of the distance to the GB plane
for different GBs.} \label{hopping}
\end{figure}

\begin{figure}
\begin{center}
\includegraphics[width=0.5\textwidth]{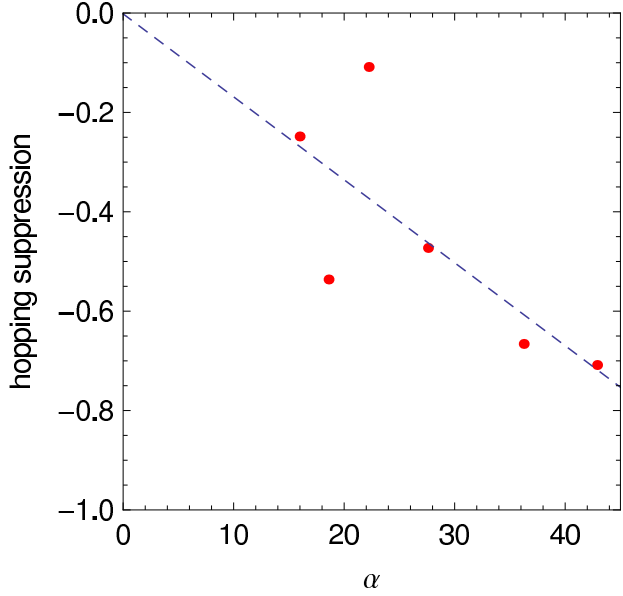}
\end{center}
\caption{The suppression of the hopping as a function of the misalignment
angle $\alpha$ (red points) and a linear fit (blue dashed line).
The hopping suppression is defined as the intgral
over a Gaussian fit of the hopping profiles shown in Fig.~\ref{hopping}.} \label{hopping1}
\end{figure}

In Fig.~\ref{hopping} we show the averaged hopping as a function
of the distance to the grain boundary plane for different GBs. If we
fit the suppressed hopping values in the vicinity of the grain boundary by
a Gaussian form and integrate over the  ``effective barrier'' derived in this way,
we find only a linear variation with misalignment angle (see Fig.~\ref{hopping1}).

\subsection{The bond valence analysis}

The structural imperfection at the grain boundary will necessarily
lead to charge inhomogeneities that will contribute --- in a similar way
as the reduced hopping --- to the effective barrier that blocks the
superconducting current over the GB. We can include these charge inhomogeneities
in our calculations by ``translating'' them into on-site potentials
on the Cu sites. It is evident that we also have to include
the charges on the O sites in our considerations although the
O sites themselves have already been integrated out in the effective one-band
tight-binding model. We will start our considerations by assigning
every atom of the perfect crystal a formal integer ionic charge:
Y$^{3+}$, Ba$^{2+}$, O$^{2-}$. The requirement of charge neutrality
leaves us with 7 positive charges to be distributed on the 3 Cu atoms:
Thus we will have two Cu$^{2+}$, and one Cu$^{3+}$ ion per unit cell.
For a crystal with bonds that are neither strictly covalent nor strictly
ionic it is convenient to introduce a fractional valence for each ion, that
is determined with respect to its ionic environment in the unit cell
of the crystal. Here we calculate the bond valence of a cation by
\[
V_i = \sum_j \exp\left(\frac{r_0 - r_{ij}}{B} \right),
\]
where the sum is over all neighboring anions, in our case the neighboring
negatively charged O and $r_{ij}$ is the cation-anion distance. 
Here we take $B=0.37$~\AA~ following
Ref.~\onlinecite{Brown}, while $r_0$ is different for
all cation-anion pairs and can also depend on the formal integer oxidation
state of the ion. The basic idea is that the bond valence sum should
agree with the assumed integer oxidation state
of the ion, and strong deviations indicate strain in the crystal or even
an incorrect structure. This seems to be a very clear concept for
the case of the Y$^{3+}$ and Ba$^{2+}$ ions. For the copper atoms,
where we can have more than one formal integer oxidation state, the
situation is slightly more complicated. Following Refs.~\onlinecite{Brown,Chmaissem}
we define $\xi_i^{(3)}$ as the fraction of Cu ions at site $i$ that are in a
Cu$^{3+}$-type oxidation state while the remaining $(1-\xi_i^{(3)})$
Cu ions are in a Cu$^{2+}$-type oxidation state. With this the average oxidation
state of the Cu ion at site $i$ is
\[
\bar{V}_i = 3\xi_i^{(3)} + 2 (1-\xi_i^{(3)})=2 + \xi_i^{(3)}.
\]
On the other hand, this should be equal to the sum of a fraction of
bond valences $V_i^{(3+)}$ of ions characterized by
$r_0(\textrm{Cu}^{3+})=1.73$~\AA~ and a fraction of bond valences
$V_i^{(2+)}$ of ions characterized by $r_0(\textrm{Cu}^{2+})=1.679$~\AA:
\[
\bar{V}_i = \xi_i^{(3)}V_i^{(3+)} + (1-\xi_i^{(3)})V_i^{(2+)}.
\]
Solving this set of equations for $\xi_i^{(3)}$ allows us to determine the fraction
of Cu ions with the valence $3+$:
\[
\xi_i^{(3)}=\frac{V_i^{(2+)} - 2}{V_i^{(2+)}-V_i^{(3+)} + 1}.
\]
In a similar way we can proceed assuming that we have a Cu ion that
could be in a 1+ or a 2+ oxidation state. Here we will use
$r_0(\textrm{Cu}^{2+})=1.679$~\AA~ and $r_0(\textrm{Cu}^{1+})=1.6$~\AA~
and we find
\[
\xi_i^{(1)}=\frac{V_i^{(2+)}-2}{V_i^{(2+)}-V_i^{(1+)}-1}.
\]
In the case that Cu$^{2+}$ and Cu$^{3+}$ are
the most probable oxidation states we will find that $\xi_i^{(3)}$
is positive and $\xi_i^{(1)}$ is negative while in the case
that Cu$^{2+}$ and Cu$^{+}$ are
the most probable oxidation states we will find that $\xi_i^{(1)}$
is positive and $\xi_i^{(3)}$ is negative. For the calculation of the
final oxidation state of a particular Cu atom one has to
use the correct, positive $\xi_i$.

In a last step we determine the fractional valence
of the O ions by summing up all charge contributions from the
neighboring cations. Here we assume that the Ba and the Y ions
are in their formal integer oxidations state
whereas we assign every Cu ion a fractional oxidation
state determined by $\xi_i^{(1/3)}$. This method ensures
that we end up with a charge neutral crystal.

In the bulk system we derive fractional valences
close to the formal integer oxidation states.
In the vicinity of the GB we find however deviations from
the bulk values due to missing or displaced neighboring ions.

\subsection{The definition of the superconducting pairing interaction}

To model the known momentum space structure of the superconducting order parameter 
in the weak coupling description of the Bogoliubov-de Gennes 
theory, one usually defines an interaction $V_{ij}$ on the bonds
connecting two nearest neighbor Cu sites. Unfortunately, there is no
obvious way to define an analogous pairing interaction in a strongly 
disordered region of the crystal, as e.g. in the vicinity of the GB, 
since the exact microscopic origin of this interaction is not known. 
Assuming that a missing or a broken Cu-O bond destroys the underlying
pairing mechanism, we develop a method of tying 
the superconducting pairing interaction between two Cu sites 
to the hopping matrix elements connecting the two sites as well as 
to the charge imbalance between them.
Hence we define the pairing interaction on a given bond as the product of 
a dimensionless constant $V_0$, that is adjusted to reproduce the
correct modulus of the gap in the bulk, and the 
hopping parameter $t_{ij}$. In addition we impose an exponential
suppression of the pairing strength with increasing charge imbalance: 
\[
 V_{ij} = V_0 t_{ij} e^{-|Q_i-Q_j|/e}.
\]
To avoid long range contributions to the pairing we use a threshhold of
$0.2$ eV for $|t_{ij}|$, thus restricting the pairing interaction to the 
bonds between nearest neighbor Cu sites in the bulk. Here we emphasize
that although we tried to model the pairing interaction in a realistic way, 
the exact procedure how we define the pairing interaction in the disordered
region does not qualitatively change the results.

\section{The calculation of the critical current}

\begin{figure}
\begin{center}
\includegraphics[width=0.7\textwidth]{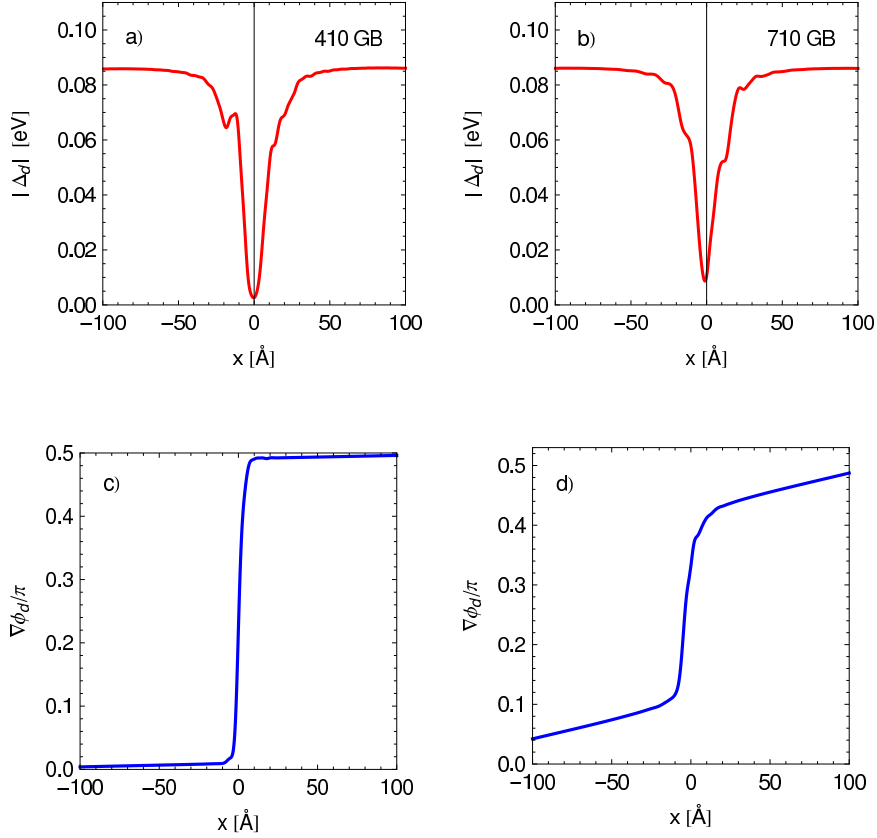}
\end{center}
\caption{The modulus (a,b) and the phase (c,d) of the
self-consistently calculated order parameter
as a function of the distance to the GB plane for a 410
and for a 710 GB.} \label{orderparam}
\end{figure}

The superconducting current over a weak link,
as well as over a metallic or an insulating barrier
is accompanied by a drastic change of the phase
of the superconducting order parameter.
For a true tunnel junction, for which the two superconducting regions
are completely decoupled, the current-phase relation is known to be sinusoidal,
and the maximum current is found for a phase jump of $\pi/2$.
For a weak link, as e.g. provided by a geometrical constriction or a
disordered region like a grain boundary, the phase of the superconducting order
parameter has to change continously between its two bulk values in the
leads on both sides of the weak link. Here one would expect deviations from the sinusoidal
current-phase relation and a smooth transition of the phase over a length scale given
by the dimensions of the weak link.
Besides the change in the phase of the order parameter in the presence
of an applied current, a weak link is also characterized by a suppression
of the modulus of the order parameter, either due to geometric restrictions or
due to strong disorder. An additional suppression of the order parameter
on the length scale of the coherence length can occur due to the
formation of Andreev bound states present at specifically orientated
interfaces of $d$-wave superconductors. Here we have calculated the order
parameter self-consistently from Eq. 8 in the main text and
found a suppression of its modulus in the vicinity of the grain boundary
that depends in its width and depth only weakly on the misalignment
angle (compare Fig.~\ref{orderparam} a,b).

\begin{figure}
\begin{center}
\includegraphics[width=0.9\textwidth]{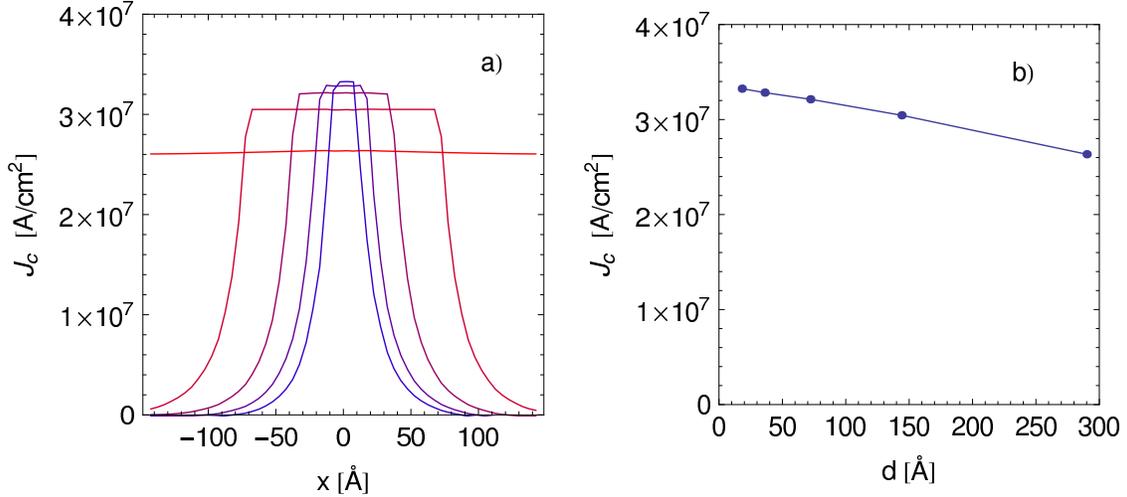}
\end{center}
\caption{(a) The current profile as a function of the distance $x$
to the GB plane for a 710 GB. The phase of the
order parameter has been fixed at different distances
$d/2$ from the GB. (b) The dependence of the critical current
as a function of the distance $d$ between the two points
at which the phase of the order parameter has been fixed.} \label{current1}
\end{figure}

For a numerical determination of the critical current ---
the maximum current that can be applied to a system without destroying its
superconducting properties --- it is convenient to enforce a certain phase
difference of the order parameter between two points of the
system separated by a distance $d$, e.g. in our example on both sides of the grain boundary region.
Here we calculate the superconducting current over the grain boundary
modelled by the Hamiltonian in Eq. 5 of the main text using
Eqs. 9 and 10 therein and fixing the phase of the order parameter
in the leads in every iteration step.
The critical current for our system can then be found as
the current maximum under a variation of the phase difference.
If a strong perturbation limits the superconducting current ---
in the so-called tunneling limit --- the main drop of the
superconducting phase will appear in the small region
of the perturbation and the change of the phase in the superconducting
leads is negligible (see Fig.~\ref{orderparam} c).
However, if the perturbation is weak, the change of the phase in the
superconducting leads becomes more important (see Fig.~\ref{orderparam} d)
and the current will depend on the distance $d$ between the two points,
at which the phase of the order parameter is fixed.
Since we are interested in the calculation of the critical current,
we have to decrease the distance, thus increasing the phase gradient,
until we reach the maximum current.
To calculate the current over the low angle grain boundaries, we have
determined the maximum current by extrapolating it to the value
expected for $d=0$ (see Fig.~\ref{current1}).


\begin{thebibliography}{00}

\bibitem{Mannhart}  Hilgenkamp, H. \& Mannhart, J. 
Grain boundaries in high-$T_c$ superconductors. 
{\it Rev. Mod. Phys.} {\bf 74}, 485 (2002).

\bibitem{Dimosetal88}
Dimos, D., Chaudhari, P., Mannhart, J. \& LeGoues, F. K. 
Orientation dependence of grain-boundary critical currents 
in YBa$_2$Cu$_3$O$_{7-\delta}$ bicrystals. 
{\it Phys. Rev. Lett.} {\bf 61}, 219 (1988).

\bibitem{Chaudhari}
Chaudhari, P., Dimos, D. \& Mannhart, J. 
Critical Currents in Single-Crystal and Bicrystal Films. 
in {\it Earlier and Recent Aspects of Superconductivity} 
(eds Bednorz, J. G. \& M\"uller, K. A.) 201-207 
(Springer-Verlag, 1990).

\bibitem{SigristRice} Sigrist, M. \& Rice, T. M. 
Paramagnetic Effect in High  $T_c$ Superconductors -
A Hint for $d$-Wave Superconductivity.
{\it J. Phys. Soc. Jpn.} {\bf 61}, 4283 (1992); 
{\it J. Low Temp Phys.} {\bf 95}, 389 (1994).

\bibitem{Gurevich} Gurevich, A., \& Pashitskii, E. A. 
Current transport through low-angle grain boundaries in 
high-temperature superconductors. 
{\it Phys. Rev. B.} {\bf 57}, 13878 (1998).

\bibitem{Stolbov} Stolbov, S. V., Mironova, M. K. \& Salama, K.
Microscopic origins of the grain boundary effect 
on the critical current in superconducting copper oxides.
{\it Supercond. Sci. Technol.} {\bf 12}, 1071 (1999).

\bibitem{Pennycook} Pennycook, S. J. {\it et al.}
The Relationship Between Grain Boundary
Structure and Current Transport in High-Tc Superconductors.
in {\it Studies of High Temperature Superconductors: 
Microstructures and Related Studies of High Temperature 
Superconductors-II, Vol. 30} (ed Narlikar, A. V.) 
Ch. 6 (Nova Science Publishers, 2000).

\bibitem{Andersen_AF_SC} Andersen, B. M., Barash, Yu. S.,
Graser, S. \& Hirschfeld, P. J. 
Josephson effects in $d$-wave superconductor junctions with
antiferromagnetic interlayers. 
{\it Phys. Rev. B} {\bf 77}, 054501 (2008).

\bibitem{Zhang} Zhang, X. \& Catlow, C. R. A. 
Molecular dynamics study of oxygen diffusion in 
YBa$_{2}$Cu$_{3}$O$_{6.91}$. 
{\it Phys. Rev. B} {\bf 46}, 457 (1992).

\bibitem{Phillpot} Phillpot, S. R. \& Rickman, J. M. 
Simulated quenching to the zero-temperature limit of the grand-canonical
ensemble. {\it J. Chem. Phys.} {\bf 97}, 2651 (1992).

\bibitem{gbangles} In the following we will call a GB with $N_{1}=1$, $N_{2}=4$ and
therefore a GB angle of $\alpha=2 \cdot 0.24074\mathrm{\,\,
rad}=27.58^\circ$ a symmetric (410) GB.

\bibitem{SlaterKoster}
Slater, J. C.  \& Koster, G. F. 
Simplified LCAO Method for the Periodic Potential Problem. 
{\it Phys. Rev.} {\bf 94}, 1498 (1954).

\bibitem{Harrison} Harrison, W. A. 
Electronic structure and the properties of solids.
(Dover Publications, 1989).

\bibitem{Liu} Liu, P. \& Wang, Y. 
Theoretical study on the structure of Cu(110)-p2$\times$1-O 
reconstruction. 
{\it J. Phys.: Condens. Matter} {\bf 12}, 3955 (2000).

\bibitem{Baetzold} Baetzold, R. C.  
Atomistic Simulation of ionic and electronic defects in 
YBa$_{2}$Cu$_{3}$O$_{7}$.
{\it Phys. Rev. B} {\bf 38}, 11304 (1988).

\bibitem{Brown} Brown, I. D. 
A Determination of the Oxidation States and Internal 
Stress in Ba$_{2}$YCu$_{3}$O$_{x}$, $x$=6-7
Using Bond Valences. 
{\it J. of Solid State Chem.} {\bf 82}, 122 (1989).

\bibitem{Chmaissem} Chmaissem, O., Eckstein, Y. \& Kuper, C. G. 
The Structure and a Bond-Valence-Sum Study of the 1-2-3
Superconductors (Ca$_{x}$La$_{1-x}$)(Ba$_{1.75-x}$La$_{0.25+x}$)Cu$_{3}$O$_{y}$
and YBa$_{2}$Cu$_{3}$O$_{y}$. 
{\it Phys. Rev. B} {\bf 63}, 174510 (2001).

\bibitem{ZhangRice} Zhang, F. C. \& Rice, T. M.
Effective Hamiltonian for the superconducting Cu oxides.
{\it Phys. Rev. B} {\bf 37}, 3759 (1988).

\bibitem{AndersenJc} Andersen, B. M., Bobkova, I.,
Barash, Yu. S. \& Hirschfeld, P. J. 
$0-\pi$ transitions in Josephson
junctions with antiferromagnetic interlayers. 
{\it Phys. Rev. Lett.} {\bf 96}, 117005 (2006).

\bibitem{Freericks} Freericks, J. K. 
Transport in Multilayered Nanostructures.
The Dynamical Mean-Field Theory Approach.
(Imperial College Press, 2006).

\bibitem{Larbalestier} Lee, S. {\it et al.}
Weak-link behavior of grain boundaries in 
superconducting Ba(Fe$_{1-x}$Co$_x$)$_2$As$_2$ bicrystals.
{\it not published}, 
available at arXiv:0907.3741.

\bibitem{Hammerl} Hammerl, G. {\it et al.}
Possible solution of the grain-boundary problem 
for applications of high-$T_c$ superconductors. 
{\it Appl. Phys. Lett.} {\bf  81}, 3209 (2002).

\bibitem{Schuster} Schwingenschl\"ogl U. \& Schuster C.
Quantitative calculations of charge-carrier densities in the depletion
layers at YBa$_2$Cu$_3$O$_{7−\delta}$ interfaces.
{\it Phys. Rev. B} {\bf 79}, 092505 (2009).

\bibitem{Alloul} Alloul, H., Bobroff, J., Gabay, M.,
Hirschfeld, P. J.
Defects in correlated metals and superconductors.
{\it Rev. Mod. Phys.} {\bf 81}, 45 (2009).

\end{thebibliography}
\end{document}